\documentclass[sigconf]{acmart}
\usepackage{fontawesome5}  

\AtBeginDocument{%
  }

\copyrightyear{2025}
\acmYear{2025}
\setcopyright{rightsretained}
\acmConference[CHIWORK '25 Adjunct]{CHIWORK '25 Adjunct: Adjunct Proceedings of the 4th Annual Symposium on Human-Computer Interaction for Work}{June 23--25, 2025}{Amsterdam, Netherlands}
\acmBooktitle{CHIWORK '25 Adjunct: Adjunct Proceedings of the 4th Annual Symposium on Human-Computer Interaction for Work (CHIWORK '25 Adjunct), June 23--25, 2025, Amsterdam, Netherlands}
\acmPrice{}
\acmDOI{10.1145/3707640.3731923  }
\acmISBN{979-8-4007-1397-2/   25/06}

\begin{document}
\title{Communicating Through Avatars in Industry 5.0: A Focus Group Study on Human-Robot Collaboration}


\author{Stina Klein}
\authornote{Both authors contributed equally to this research.}
\email{stina.klein@uni-a.de}
\orcid{0000-0003-4998-6811}
\affiliation{%
  \institution{University of Augsburg}
  \city{Augsburg}
  \country{Germany}
}

\author{Pooja Prajod}
\authornotemark[1]
\orcid{0000-0002-3168-3508}
\email{Pooja.Prajod@cwi.nl}
\affiliation{%
  \institution{Centrum Wiskunde \& Informatica}
  \city{Amsterdam}
  \country{The Netherlands}
}

\author{Katharina Weitz}
\orcid{0000-0003-1001-2278}
\email{katharina.weitz@hhi.fraunhofer.de}
\affiliation{%
  \institution{Fraunhofer Institute for Telecommunications, Heinrich Hertz Institute, HHI}
  \city{Berlin}
  \country{Germany}
}

\author{Matteo Lavit Nicora}
\orcid{0000-0003-4256-7606}
\email{matteo.lavitnicora@cnr.it}
\affiliation{
  \institution{National Research Council of Italy, Institute of Intelligent Industrial Technologies and Systems for Advanced Manufacturing}
  \city{Lecco}
  \country{Italy}
}

\author{ Dimitra Tsovaltzi}
\orcid{0000-0002-1670-5799}
\email{dimitra.tsovaltzi@dfki.de}
\affiliation{
  \institution{German Research Center for Artificial Intelligence}
  \city{Saarbrücken}
  \country{Germany}
}

\author{Elisabeth André}
\orcid{0000-0002-2367-162X}
\email{elisabeth.andre@uni-a.de}
\affiliation{%
  \institution{University of Augsburg}
  \city{Augsburg}
  \country{Germany}
}

\renewcommand{\shortauthors}{Klein and Prajod et al.}

\begin{abstract}
The integration of collaborative robots (cobots) in industrial settings raises concerns about worker well-being, particularly due to reduced social interactions. Avatars - designed to facilitate worker interactions and engagement - are promising solutions to enhance the human-robot collaboration (HRC) experience. However, real-world perspectives on avatar-supported HRC remain unexplored. To address this gap, we conducted a focus group study with employees from a German manufacturing company that uses cobots.  Before the discussion, participants engaged with a scripted, industry-like HRC demo in a lab setting. This qualitative approach provided valuable insights into the avatar's potential roles, improvements to its behavior, and practical considerations for deploying them in industrial workcells. Our findings also emphasize the importance of personalized communication and task assistance. Although our study's limitations restrict its generalizability, it serves as an initial step in recognizing the potential of adaptive, context-aware avatar interactions in real-world industrial environments.
\end{abstract}




\maketitle

\section{Introduction}
Introduced around 2020, Industry 5.0 represents a paradigm shift toward human-centric manufacturing~\citep{sharma2020evolution, xu2021industry, adel2022future, rovzanec2023human}. Unlike previous industrial revolutions that primarily focused on replacing human labor with automation, Industry 5.0 emphasizes collaboration between human workers and collaborative robots (cobots). Imagine a cobot and a worker sharing a workspace, working in close proximity and simultaneously working on a shared object – this exemplifies the collaborative workflow in Industry 5.0~\citep{prajod2024face, mondellini2023behavioral, zafar2024exploring, klein2024creating}.
A crucial concern in this collaborative setting is the potential impact of introducing cobots on workers’ mental health and well-being. Many industrial workcells, particularly in small- and medium-sized enterprises, involve a single worker interacting with a cobot, reducing opportunities for human-human interactions~\citep{hovens2020workplace, muhlemeyer2020assessment, osika2023humanistic}. This reduction in social interactions can contribute to emotional loneliness, which in turn may negatively affect employee performance~\citep{akccit2017relationship} and hinder the acceptance of cobot systems~\citep{welfare2019consider, meissner2020friend}.

Beyond social isolation, additional factors such as task complexity, imbalanced workload distribution, and cobot speed may introduce stress, frustration, and boredom, which can further reduce worker well-being~\citep{welfare2019consider, meissner2020friend, ShelloReview, prajod2024face}. Until recently, research in industrial human-robot collaboration (HRC) has prioritized the physical safety of workers~\citep{robla2017working, bragancca2019brief, arents2021human}. However, Industry 5.0 extends this focus and advocates that worker-cobots interactions are not only physically safe but also contribute positively to the workers' mental health and well-being~\citep{grosse2023human, loizaga2023comprehensive}. By integrating social interaction capabilities, cobots have the potential to act as supportive collaborators, mitigating feelings of loneliness and building a more engaging work environment.

Cobots are typically robotic arms and have limited social interaction capabilities. One approach for circumventing this limitation is leveraging avatars or virtual characters to recreate some aspects of social interaction~\cite{arora2024socially, watanabe2023augmented}. Avatars have been employed in various applications to represent social roles, such as well-being coaches~\citep{el2020virtual}, tutors~\citep{armando2022impact}, and training partners~\citep{bosman2019virtual}. In an industrial HRC scenario, an avatar can serve as a social interface for the physical cobot~\cite{beyrodt2023socially, nunnari2025socially, nicora2023towards}, making it feel more like a teammate. An avatar can also play a valuable role in communicating interventions and explaining the cobot's decisions related to adaptations and recommended actions~\cite{harbers2009study, weitz2021let, arora2022employing}.

We conducted a focus group interview involving three employees from a German manufacturing company that deploys cobots in their production line. The employees had different job profiles, representing the different perspectives of the various stakeholders. For supporting discussions, they first interacted with an industry-like human-robot collaboration scenario that is augmented by the presence of an avatar. The avatar contributed to the interaction through interventions aimed at addressing the operator's affective state and motivating them. The insights from the focus group help us identify factors that could improve the well-being of industrial workers and future research directions.

\section{Methodology}

\subsection{Setup}

\begin{figure*}
    \centering
    \includegraphics[width=0.8\textwidth]{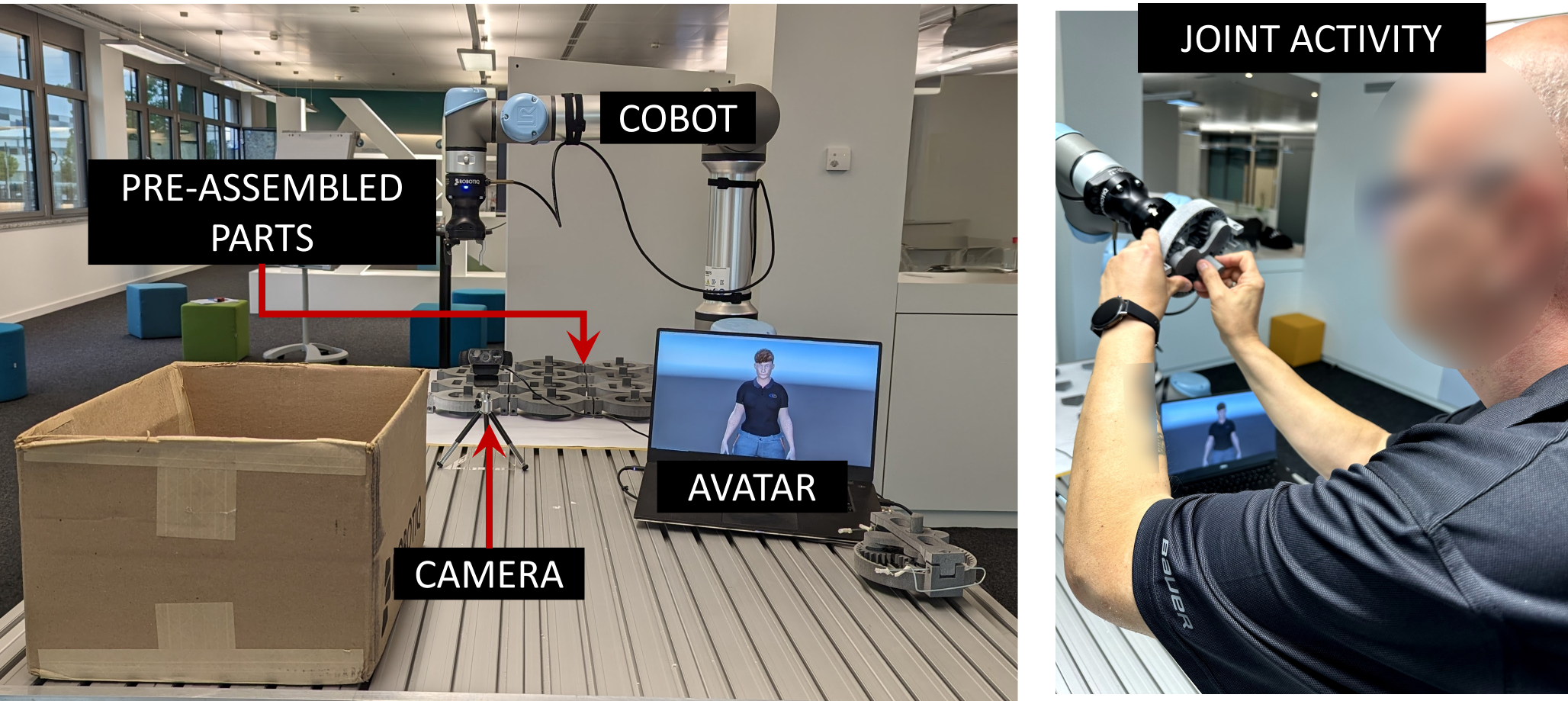}
    \caption{The image on the left shows our setup for collaborative assembly, and on the right is an example of joint activity involving the cobot and the operator.}
    \label{fig:demo_setup}
    \Description{test description for accessibility check.}
\end{figure*}

We used an established experimental setup in which the cobot's production rate was manipulated to elicit varying collaboration experiences~\citep{beyrodt2023socially, prajod2024flow, mondellini2024exploring}. We implemented a collaborative assembling task where an operator and a cobot assembled 3D-printed gearboxes. The operator assembled half of the assembly, and the other half was pre-assembled on the table for the cobot (see Figure~\ref{fig:demo_setup}). At specific intervals, the cobot picked one of the sub-assemblies and brought it to the operator. The operator meshed the sub-assemblies together in a joint activity to produce the finished product. 

We manipulated the cobot's picking interval to implement three operating modes: Slow, Fast, and Adaptive. In the Slow mode, the cobot scanned over the sub-assemblies for around one minute before bringing one for joint activity. So, the cobot was designed to be slower than the operator, resulting in the operator having to wait for the cobot. In the Fast mode, the cobot picked the sub-assembly for joint activity without any scanning motion. This mode resulted in the cobot waiting for the operator to complete their sub-assembly. The Adaptive mode was designed to simulate synchronous completions of sub-assemblies by the cobot and the operator. In this case, the cobot scanned over the sub-assemblies until the operator was ready for joint activity.

\begin{figure*}
    \centering
    \includegraphics[width=\linewidth]{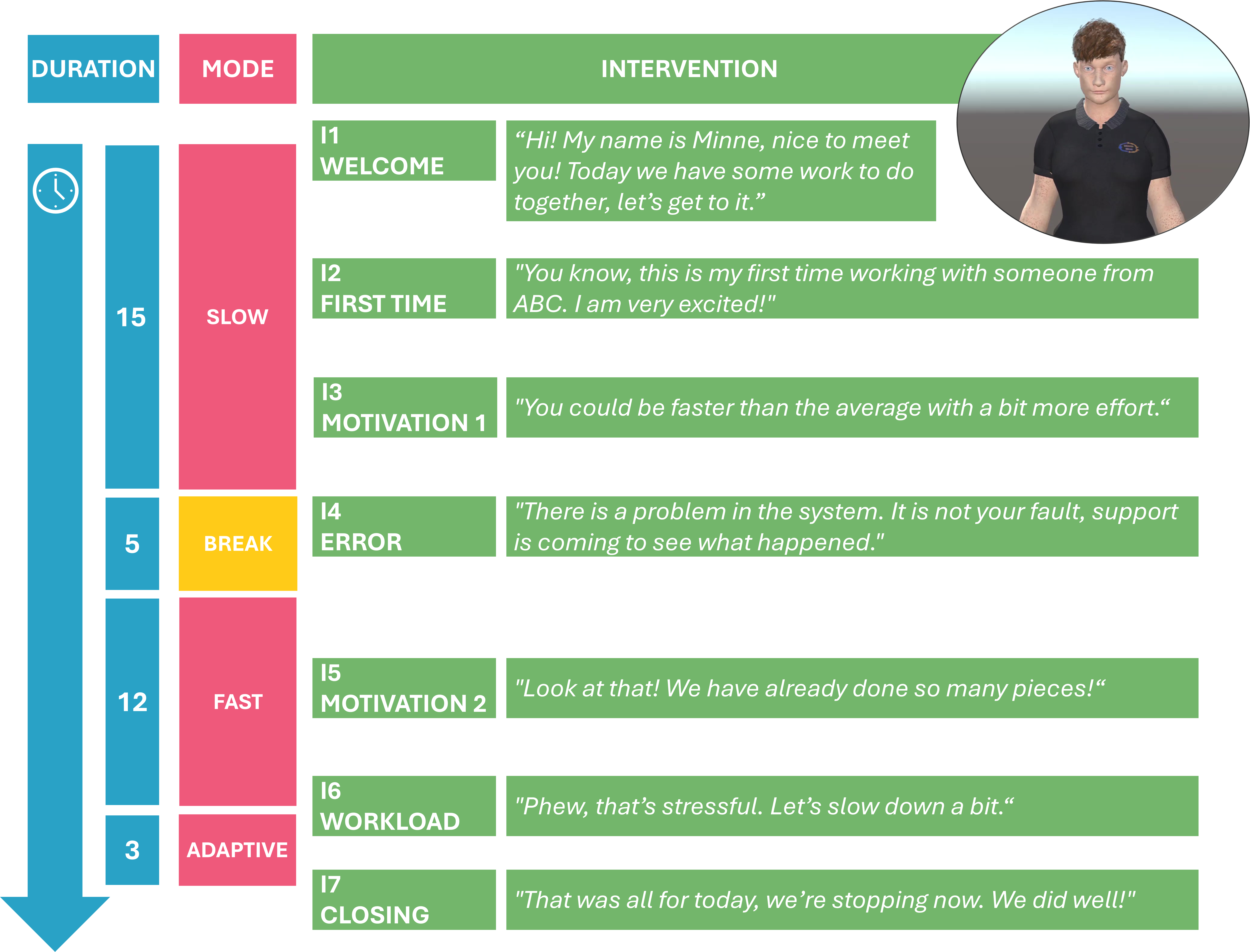}
    \caption{Flow chart of the demo scenario.}
    \label{fig:workflow}
\end{figure*}

Inspired by other works~\citep{nicora2023towards, nunnari2025socially} that investigated avatars in industrial settings, an avatar was displayed on a screen close to the cobot's base, as seen in Figure~\ref{fig:demo_setup}. It had idle animations such as breathing and blinking. A camera was placed facing the operator, which was used to create an illusion that the avatar interventions (described in Section~\ref{sec:demo_script} and depicted in Figure \ref{fig:workflow}) were based on the operator's predicted affective state. The camera input was not used in our demo scenario, as all the interactions (including the avatar interventions) were completely scripted.

\subsection{Demo Scenario}
\label{sec:demo_script}

After introducing the work cell to the participants, we first let them practice assembling some gearboxes with the cobot. Once they are familiar with the steps, the demo starts with the avatar introducing itself in the first avatar intervention (\textbf{I1}). All instructions and avatar interventions were in German. 

In the first 15 minutes of the demo, the cobot operated in Slow mode, which led to the operator waiting for the cobot to bring the sub-assembly.  
After 5 (\textbf{I2}) and again after 10 (\textbf{I3}) completed gearboxes, the avatar intervened.
The session continued until the 15-minute mark. At this point, we introduced a scripted error. The cobot came to a halt, and the avatar informed the operator that there was an error in the system (\textbf{I4)}. 
The participant was sent on a break while the experimenters acted like they were fixing the system. After around 5 minutes, the experimenters informed the participant that they could resume the demo session, but the cobot would operate at a faster pace to make up for the lost time. This marked the beginning of the Fast mode, where the cobot picked the sub-assemblies without any scanning. 
This condition lasted for around 12 minutes 
After 8 gearboxes were completed in the fast condition, the avatar intervened for the fifth time (\textbf{I5}).
For the last 3 minutes of the demo, the cobot picking rate was approximately matched to the operator's assembling speed. We refer to this as the Adaptive mode. During the transition from Fast to Adaptive mode, the avatar intervened again (\textbf{I6}).
The demo ended after a few minutes of Adaptive mode, and the avatar intervened one last time, closing the demo scenario (\textbf{I7}).

\subsection{Data Collection and Analysis}
We conducted a one-day workshop, where we invited three visitors from a German manufacturing company that already deploys cobots in the production line. During the workshop, they interacted with the above-described cobot-avatar system and participated in a focus group discussion. The participants (2 male, 1 female) had different roles in the company - a cobot worker (CW), a learning\&development manager (LDM), and an ergonomics\&health manager (EHM) - giving us perspectives of different stakeholders. 
The interview was conducted following the story interview method~\citep{mackay2023doit}. This method involves asking the participant to provide a walk-through of a recent experience (in our case, the scripted demo), followed by specific questions to delve deeper into certain aspects of the experience. Two microphones recorded the interview. Afterwards, the researchers transcribed it.
The interview data were analyzed independently by two researchers following the reflexive thematic analysis by~\citet{braun2006using}.

\section{Results}
\label{sec:results}
The interviews provided valuable insights into the perception of the robot and the avatar during the demo, as well as future ideas and limitations associated with their integration in real-world settings.
\subsection{Avatar and Robot Perception}
Overall, the \textbf{presented task} during the demo was perceived as realistic by all three participants. Nevertheless, they stated that the steps of the work processes could be optimized to improve overall efficiency. Especially the position of the \textbf{robot} could be adjusted to make it more accessible and to facilitate a smoother hand-over between human and robot. 
The role of the \textbf{avatar} (see Appendix \ref{sec:appendix_a} Table \ref{tab:robot_avatar}) was perceived differently, depending on the context of the scenario. The lack of facial reaction of the avatar was mentioned, suggesting that the verbal expressions were identified as a key factor of being aware of the avatar. The personal communication through the avatar was also perceived positively. Participants viewed the avatar in three different roles: (1) as a \textbf{supervisor}, pressuring them to work faster, (2) as a \textbf{colleague}, who was relaxed and non-disruptive, and (3) as an \textbf{impersonation of the worker}. Regarding the content of spoken communication, we found the following feedback based on  the avatar's intervention:
\begin{itemize}
    \item \textbf{I1 -- Welcome}: Participants had issues understanding the name (e.g., \textit{``What kind of name is that?'}' --- LDM) and had wished for a detailed introduction of the task by the avatar (instead of the experimenters) (e.g., \textit{``Let's get started. But on the other hand. What should we start with? You've told us what we need to do. But he [the avatar] could have done that too.''}  --- EHM)
    \item \textbf{I2 -- First Time}: The participants felt that the avatar addressed them personally by naming the company: \textit{``I thought it was nice. So it speaks to you very personally with stating the company name''} --- EHM)
    \item \textbf{I3 -- Motivation 1}: This intervention was perceived as rude (e.g., \textit{``It can be glad it was an avatar''} --- CW). This information could have been shared in a more friendly and positive way. Moreover, it would be much better if the avatar also explained what the operator could do differently to improve.
    \item \textbf{I4 -- Error }: It was seen as a neutral sentence aimed at confirming that there is a problem.  The avatar said that the support is coming, which was nice and honest (e.g., \textit{``Just such a completely neutral sentence and it is immediately offered help. I like that''} --- EHM)
    \item \textbf{I5 -- Motivation 2}: The participants perceived it as more positive than motivation sentence 1: \textit{``In comparison to the second last sentence `with a little more effort you could be better than average,' it was a refresher.''} --- CW
    \item \textbf{I6 -- Workload}: The statement was perceived as appropriate: \textit{``Okay, let's slow down. That's no problem'' } --- CW 
    \item \textbf{I7 -- Closing }: The sentence was fine but perceived as unrealistic: \textit{``Yes, it was okay, but as I said, it can't be done. ... it can't be done with the number of units.''} --- CW
\end{itemize}

\subsection{Improvements}
The participants envisioned \textbf{improvements} to the avatar functionality, particularly in areas that enhance workplace support for the worker (see Appendix \ref{sec:appendix_a} Table \ref{tab:Future}). It should ensure \textbf{speed adoption} by smoothly integrating into workflows and supporting \textbf{workload management} through break reminders and job rotations. Enhancing \textbf{personalized feedback} and taking on an \textbf{infotainment role} were also proposed. Additionally, \textbf{error assistance} and \textbf{safety compliance} could improve accuracy and adherence to company regulations. The avatar might act as a \textbf{motivator} to boost the motivation of workers and use \textbf{emotional intelligence} to adapt interactions based on workers’ moods. Future designs should balance efficiency with personalization for better user acceptance.

\subsection{Real-world Limitations}
Despite the potential benefits, participants identified several \textbf{limitations} that must be considered when implementing avatars in real-world industry settings. These limitations can be categorized into three main areas: \textbf{real-world constraints}, \textbf{speech clarity}, and \textbf{privacy concerns}.
First, \textbf{real-world limitations} highlight the fact that workplace settings impose constraints that have to be taken into account (see Appendix \ref{sec:appendix_a} Table \ref{tab:Limits}). Employees have to fulfill the realities of their work schedules and production demands.
The cobot worker illustrated this by stating: \textit{``I can't say I'm doing 300 engines today, and then I am done. I can't say ‘I'm going home now.’ Even if he [\textit{the avatar}] says ‘It's time to go home’ 20 times. It can't be realized that way.''}
Second, \textbf{speech clarity} emerged as a concern, particularly in environments where multiple avatars might be used simultaneously. In certain workplaces, the avatar's voice could be challenging to understand due to background noise or overlapping audio cues. The manager expressed skepticism by stating: \textit{``It definitely wouldn't work in every workplace.''  --- EHM }
While the worker added regarding multiple avatars:
\textit{``That will make me very nervous.''}
Finally, \textbf{privacy concerns} were a substantial issue, as both managers expressed strong opposition to data recording and information sharing. This was captured in the statement by LDM:
\textit{``That would be a no-go, for sure. So recording or passing on information, we don't even want to discuss that, that wouldn't work.''}
Their agreement emphasized the widespread discomfort with potential surveillance or data misuse in the workplace.

Overall, these findings suggest that while the avatar system may offer benefits, its implementation must carefully consider workplace constraints, communication challenges, and privacy expectations to ensure acceptance and usability.

\section{Discussion and Conclusion}

With this focus group study, we aimed to explore factors that support workers' well-being in avatar-supported HRC in an Industry 5.0 setting. Overall, the results indicate that the avatar was perceived positively, attributing roles ranging from supervisor to colleague. The avatar interventions were generally seen as beneficial, though the content and tone of the avatar intervention should avoid pressuring or criticizing the worker. The results fit nicely into the current literature, going beyond safety concerns~\citep{robla2017working, bragancca2019brief, arents2021human}, and addressing the well-being of workers~\citep{grosse2023human, loizaga2023comprehensive}.   
In that line, we can infer several lessons learned for improving the avatar in HRC and, subsequently, improving the worker's well-being: 

\begin{enumerate}
    \item \textbf{Personalized communication} substantially impacted the perception of the avatar, emphasizing the importance of context-aware and emotionally intelligent interaction designs.
    \item \textbf{Practical improvements}, such as assistance in case of an error and compliance with safety regulations, could further enhance productivity and sustain company regulations.
    \item \textbf{Real-world circumstances}, like privacy concerns, environmental considerations such as noise, and workload, must be considered for a favorable design of avatars in HRC.  
\end{enumerate}
A key contribution of our research is the combination of feedback from both a worker and management from a company deploying cobots, together with an experimental setup testing the avatar during a demo.
However, the study is not without limitations. The scripted and controlled interaction setting lacked realism, probably affecting the responses given by participants. Additionally, while the focus group study allowed for an in-depth discussion of the avatar considering different perspectives, it only considered stakeholders of one company, limiting the generalizability.
Accordingly, we can identify several research opportunities for the future: the enhancement of the current avatar by integrating adaptive personalization aspects, but also a stronger emphasis on real-world settings, and a focus on explainability to improve collaboration ~\citep{hald2021error}. Here, the combination of different methods, such as co-creation workshops~\citep{weitz2024explaining} and testing of a fully functional prototype in the wild, can extend our findings and address the limitations.
Our study offers first insights into positive perceptions of the avatar, it remains an open question whether such systems can genuinely substitute for human social interactions or merely complement them in specific contexts. This opens even more opportunities for future research. 
In conclusion, avatars in HRC offer promising possibilities to support workers' well-being. Their integration must address real-world constraints and should focus more on personalization.

\begin{acks}
    The authors gratefully acknowledge partial funding of this work by the Bavarian Research Foundation within the project FORSocialRobots (AZ-1594-23).
\end{acks}

\bibliographystyle{ACM-Reference-Format}
\bibliography{sample-base}

\appendix
\section{} 
\label{sec:appendix_a}

This appendix contains Table 1-3, which give an overview of the categories as well as quotes from the interview summarized in Section \ref{sec:results}.

\begin{table*}
    \caption{Interview insights on robot and avatar perception.}
    \label{tab:robot_avatar}
    \renewcommand{\arraystretch}{0.9} 
    \centering
    \begin{tabular}{|c|p{2.5cm}|p{3.5cm}|p{9cm}|}
        \hline
        \textbf{Category} & \textbf{Subcategory} & \textbf{Perception} & \textbf{Example Quote} \\
        \hline
        \faRobot & Improvements & The robot's actions could be optimized to improve the working situation. & \textit{``Perhaps the position of the robot would have been a bit better if I had positioned it a bit in the opposite direction instead of at such an angle that the sprockets would have fallen into it. ... Then you wouldn't have had to be so careful about holding on.''} --- CW \\
        \hline
        \faUser & Perception & Missing Facial Reaction & \textit{``So the avatar was always very neutral when I looked at it.''} --- CW \\
        & & Verbal & \textit{``So the first time I really noticed him was when he said: ‘I like working.’''} --- EHM \\
        & & Personal communication & \textit{``I thought that was nice.''} --- CW \\
        \hline
        \faUser & Role & Boss & \textit{``It was more like the master standing beside you. Hurry up.''} --- CW \\
        & & Colleague & \textit{``In this context, I have seen the robot twice as my colleague in front of me'' ``He doesn't talk back. He doesn't grumble, doesn't drink coffee, is wonderful, doesn't need a smoke break, doesn't go to the toilet. It couldn't be better.''} --- CW \\
        & & Impersonation & \textit{``the avatar should in the end represent me.''} --- CW \\
        \hline
        \multicolumn{4}{|p{\dimexpr\textwidth-2\tabcolsep}|}{\textbf{Symbols: } \faRobot : Robot \quad \faUser : Avatar} \\
        \hline
    \end{tabular}
\end{table*}

\begin{table*}
    \caption{Interview Insights of participants on future ideas and improvements.}
    \label{tab:Future}
    \renewcommand{\arraystretch}{0.95}
    \centering
    \begin{tabular}{|p{2cm}|p{4.5cm}|p{10cm}|}
        \hline
        \textbf{Category} & \textbf{Description} & \textbf{Example Quote} \\
        \hline
        Speed Adoption & The avatar cannot stop overall processes but have to be integrated into them. & \textit{``We have a fixed cycle time. This means that we cannot stop the cobot when the avatar or the system notices, ..., the employee is getting tired, the employee is no longer focused.''} --- EHM \\
        \hline
        Workload Management & The avatar could suggest breaks or adjustments to reduce monotony. & \textit{``The avatar could tell someone else ‘The employee is getting tired, we need to do a job rotation’ or to tell the employee themselves: ‘You need to take a short break’''} --- EHM \\
        \hline
        Personalized Feedback & The avatar could provide personalized feedback based on individual work habits. & \textit{``If you have a system that tells you how you are feeling physically.''} --- EHM \\
        \hline
        Infotainment Role & The avatar could provide information based on the workers preferences. & \textit{``Hey [\textit{name of the worker}], the road is clear, you can relax and there's no traffic jam on your way home. You can go home relaxed.``That is the menu in the canteen today.''} --- LDM \\
        \hline
        Error Assistance & The avatar could help identify and correct errors early in the process. & \textit{``Yes, so I think it would be great ... if one of the work steps is carried out incorrectly, he says that ahead of time, that a work step has been carried out incorrectly or not, then put someone back in the right direction. That would be cool.''} --- EHM \\
        \hline
        Motivator & The avatar could engage in casual conversations to improve workplace atmosphere. & \textit{``In practice, you could say it's the end of the day now. We reached the quantity today and the quality was also 85,000.''} --- LDM \\
        \hline
        Safety Compliance & The avatar could remind workers to follow safety protocols or refresh the content of the safety protocols. & \textit{``There are already many, many documents, many work instructions that we have at the workplaces, i.e. how a job is to be carried out, where it says what safety equipment the employee must have, what gloves they must wear. ... But it would be much, much, much more personal if there was an avatar and would explain it to you and then maybe again.''} --- EHM \\
        & & \textit{``Would I rather hear it from the avatar that I'm wearing the wrong gloves? Or do I hear it from some superior? And I think everyone prefers to hear it from a computer''} --- EHM \\
        \hline
        Emotional Intelligence & The avatar could adapt to workers’ moods and provide context-aware assistance. & \textit{``If you say: okay, we have an avatar there who greets you personally, who then perhaps quickly realizes how you are in a good mood in the morning or how attentive you are and then perhaps explains the most important things to you again, just like that.''} --- EHM \\
        \hline
    \end{tabular}
\end{table*}

\begin{table*}
    \caption{Limits stated by participants.}
    \label{tab:Limits}
    \renewcommand{\arraystretch}{1}
    \centering
    \begin{tabular}{|p{2.5cm}|p{6.5cm}|p{7.5cm}|}
        \hline
        \textbf{Category} & \textbf{Description} & \textbf{Example Quote} \\
        \hline
        Real World Limitations & Work settings have to be taken into account. & 
        \textit{``I can't say I'm doing 300 engines today, and then I am done. I can't say ‘I'm going home now.’ Even if he [\textit{the avatar}] says ‘It's time to go home’ 20 times. It can't be realised that way.''} --- CW \\
        \hline
        Speech Clarity & Depending on the working environment, the avatar's voice might be difficult to understand, especially if several agents are used simultaneously. & 
        \textit{``It definitely wouldn't work in every workplace.''} --- EHM 
        
        \textit{``That will make me very nervous.''} --- CW \\
        \hline
        Privacy & Data recording and information sharing is not wanted. & 
        \textit{``That would be a no-go, for sure. So recording or passing on information, we don't even want to discuss that, that wouldn't work.''} --- LDM \\
        \hline
    \end{tabular}
\end{table*}


\end{document}